\documentclass{ws-ijmpe}

\usepackage{epsfig}

\begin{document}

\markboth{Thorsten Renk}{YaJEM --- a Monte Carlo code for in-medium shower evolution}

\catchline{}{}{}{}{}

\title{YaJEM --- a Monte Carlo code for in-medium shower evolution}

\author{\footnotesize Thorsten Renk}

\address{Department of Physics, P.O. Box 35, FI-40014 University of Jyv\"{a}skyl\"{a}, Finland and\\
Helsinki Institute of Physics, P.O. Box 64, FI-00014, University of Helsinki, Finland
\\
thorsten.i.renk@jyu.fi}

\maketitle

\begin{history}
\received{(received date)}
\revised{(revised date)}
\end{history}

\begin{abstract}
High transverse momentum ($P_T$) QCD scattering processes are regarded as a valuable tool to study the medium produced in heavy-ion collisions, as due to uncertainty arguments their cross section should be calculable independent of medium properties whereas the medium  then modifies only the final state partons emerging from a hard vertex. With the heavy-ion physics program at the CERN LHC imminent, the attention of high $P_T$ physics in heavy ion collisions is shifting from the observation of hard single hadrons to fully reconstructed jets. However, the presence of a background medium at low $P_T$ complicates jet-finding as compared to p-p collisions. Monte-Carlo (MC) codes designed to simulate the evolution of parton showers evolving into hadron jets are valuable tools to understand the complicated interplay between the medium modification of the jet and the bias introduced by a specific jet-finding scheme. However, such codes also use a set of approximations which needs to be tested against the better understood single high $P_T$ hadron observables. In this paper, I review the ideas underlying the MC code YaJEM (Yet another Jet Energy-loss Model) and present some of the results obtained with the code.
\end{abstract}

\section{The concept of jets in medium}

Jets in vacuum are best understood as a tool to bridge the gap between theory and experiment: While using perturbative Quantum Chromodynamics (pQCD) one can calculate results in terms of high $p_T$ parton production cross sections, experimentally collimated showers of hadrons are observed. In nature, the link between these is given by the perturbative parton shower evolution which turns initial high virtuality partons by braching processes into a shower of softer, lower virtuality partons, and by the non-perturbative hadronization of this parton shower. Jets are then an operational definition how to treat the observed hadron distribution to 'undo' this evolution and combine properties of measured hadrons into an object, the jet, which can be compared with calculations on the parton level. For instance, sequential recombination algorithms like anti-$k_T$ try to undo the branchings of the shower evolution by combining pairs of hadrons close in phase space. This means that a jet is only a meaningful concept in the context of a particular jet finding algorithm, and that the choice of the algorithm always corresponds to a bias to pick out a particular subset of all possible evolutions of the original parton.

In the presence of the soft medium created in heavy-ion collisions, the picture becomes more complicated. First, even if a jet which is completely uncorrelated with the medium is embedded into the background, the results of jet finding may change. This is because the medium constitutes a noisy environment in which the soft hadrons of the jet cannot be uniquely identified, thus a medium hadron may be combined into part of the jet or a jet hadron may be regarded as part of the background. Such complications can be studied by embedding MC jets into a medium background and trying to recover them.

However, in general the role of the medium is much more pronounced. There is expected to be significant interaction between the parton shower and the medium, i.e. there can be substantial redistribution of energy and momentum from the perturbatively calculable part of the system to the non-perturbative background and vice versa: The medium can 'absorb' partons which become sufficiently soft, whereas hard scatterings between shower and medium can kick a parton from the medium and correlate it with the jet.
This has two important consequences: First, the connection of any jet definition at 'hadron level' with the underlying parton becomes tenuous, as a medium hadron which did not originate in the branching process of the initial parton may still carry part of its energy and momentum picked up via elastic collisions. Instead, the properties of the initial parton are only manifest in the flow of energy and momentum carried by both perturbative and non-perturbative sector. Second, there is good evidence that the redistribution of energy and momentum in the soft bulk is quite different from the dynamics of a parton shower --- hydrodynamical phenomena such as shockwaves and diffusion wakes turn out to be important \cite{Betz,Bryon}. Consequently, vacuum jet definitions designed to unfold perturbative phenomena may largely miss this part of the evolution.

There are then two possible classes of jet definitions suitable for the heavy-ion environment. The first class focuses on the perturbative part of the evolution and eliminates the non-perturbative physics by cuts e.g. on $P_T$. The medium effect is then only apparent from the functional behaviour of jet $R_{AA}$ as a function of the cuts as suggested e.g. in \cite{Vitev}. The second class adapts to the non-perturbative physics and tries to capture as much as possible of the initial parton energy and momentum. This may require jet definitions very different from the vacuum case.

With this in mind, some care has to be taken to interpret the results of in-medium shower evolution codes correctly. Such codes focus on the {\em perturbative} side of the evolution only. This means that the code will either make the somewhat artificial assumption that no energy and momentum are exchanged between medium and parton (as e.g. in Q-PYTHIA \cite{QPYTHIA}) or that the energy in the parton shower will not balance the energy of the initial hard parton because the medium energy balance is not explicitly included.

\section{A description of YaJEM}

YaJEM is an example for an in-medium shower code which permits the exchange of energy and momentum between medium and shower, but does not explicitly model the medium. The model is described in detail in \cite{YaJEM1,YaJEM2,YaJEM3}. It is based on the PYSHOW algorithm \cite{PYSHOW} which models the shower as a series of $1\rightarrow2 $ branchings of partons $a \rightarrow bc$ and to which it reduces in the absence of a medium. The equations resulting from pQCD expressions for the branching probabilities are solved by a MC in momentum space.
In a medium, the medium evolution in position space must be connected with the shower evolution equations in momentum space. Within YaJEM the link to the medium spacetime dynamics is made by modelling the average time for a parton $b$ to branch from parent $a$ given the parton energies and virtualities based on the uncertainty relation as

\begin{equation}
\label{E-time}
\langle \tau_b  \rangle= \frac{E_b}{Q_b^2} - \frac{E_b}{Q_a^2}
\end{equation}  

whereas the actual time in given branching is generated from the exponential branching probability distribution

\begin{equation}
\label{E-time-r}
P(\tau_b) = \exp\left[- \frac{\tau_b}{\langle \tau_b \rangle}  \right].
\end{equation}

For simplicity, all partons of a shower are propagated with this time information along an eikonal trajectory determined by the shower initiator. This amounts to neglecting the spread in transverse space when probing the medium. Currently, YaJEM models three different scenarios for the parton-medium interaction, two of which modify the kinematics of the propagating parton whereas the last modifies the branching probabilities at each vertex (thus, in the first two scenarios the energy of the shower is {\em not} the energy of the shower initiating parton as there is explicit energy transfer between shower and medium, whereas in the last scenario the energy in the shower is conserved). For instance, in the RAD scenario, the medium is assumed to cause an increase $\Delta Q_a^2$ in the virtuality of a parton $a$ based on a local transport coefficient $\hat{q}(\zeta)$ as given by the line integral

\begin{equation}
\Delta Q_a^2 = \int_{\tau_a^0}^{\tau_a^0 + \tau_a} d\zeta \hat{q}(\zeta).
\end{equation}

This modification causes medium-induced radiation. In the DRAG scenario the medium is assumed to cause energy and momentum loss along the parton trajectories, whereas in the FMED scenario the kinematics remains unmodified, but the branching probabilities are modified as compared to the vacuum case. In general, in the RAD scenario the energy in the final perturbative state is formally larger than initially (the depletion of medium partons becoming corelated with the jet is neglected) whereas in the DRAG scenario the final energy is smaller (the absorption of partons by the medium us neglected). These effects in principle have to be compensated for when embedding YaJEM jets into a background.

\section{Key YaJEM results}

\begin{figure}[htb]
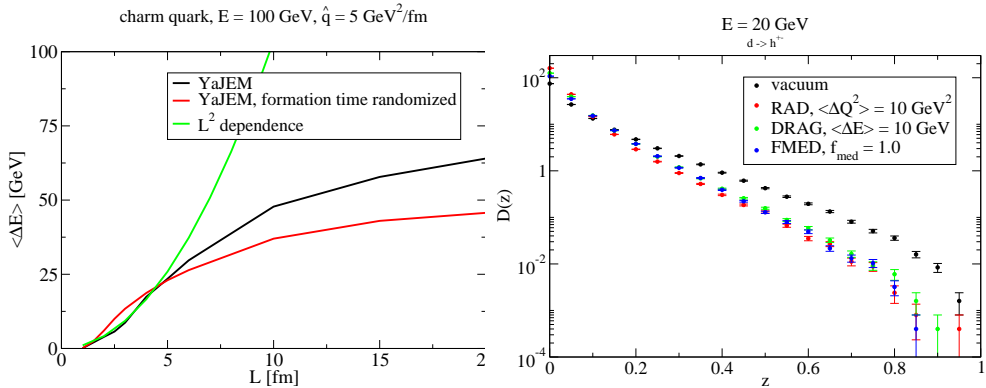

\epsfig{file=dEdx, width=6.5cm}{\epsfig{file=D_comp.eps, width=6.5cm}}
\caption{\label{F-YaJEM1} Left panel: Mean energy loss per unit pathlength from a 100 GeV charm quark propagating through a constant medium for two different assumptions about the spacetime picture of the shower, compared with an $L^2$ dependence. Right panel: Examples of medium-modified fragmentation functions obtained in YaJEM for three different scenarios of parton-medium interaction.}
\end{figure}

Before applying any in-medium shower evolution to jets, it is useful to understand how it relates to known models and observables of leading parton energy loss. An important question is to what degree the LPM suppression  of subsequent induced radiation processes known to lead to the characteristic $L^2$ dependence of energy loss in a constant medium is preserved in a probabilistic MC description. This has been computed for YaJEM in \cite{YaJEM2} and is shown in Fig.~\ref{F-YaJEM1} left for a 100 GeV charm quark propagating through a constant medium in the RAD scenario.

If only Eq.~(\ref{E-time}) is used to model the spacetime picture of the shower based on the \emph{average} lifetime of a virtual state, then there is a region in which $L^2$ dependence is seen before finite energy corrections become manifest. However, if  Eq.~(\ref{E-time-r}) is used to determine the lifetime, this region is substantially reduced.  Thus, while the MC code is capable of preserving this important feature known from analytic calculations, an attempt to model more realistically much weakens it. This has also been observed in a different framework \cite{JEWEL}. The disagreement with the pathlength dependence deduced from data is currently an unsolved theoretical issue.

\begin{figure}[htb]
\epsfig{file=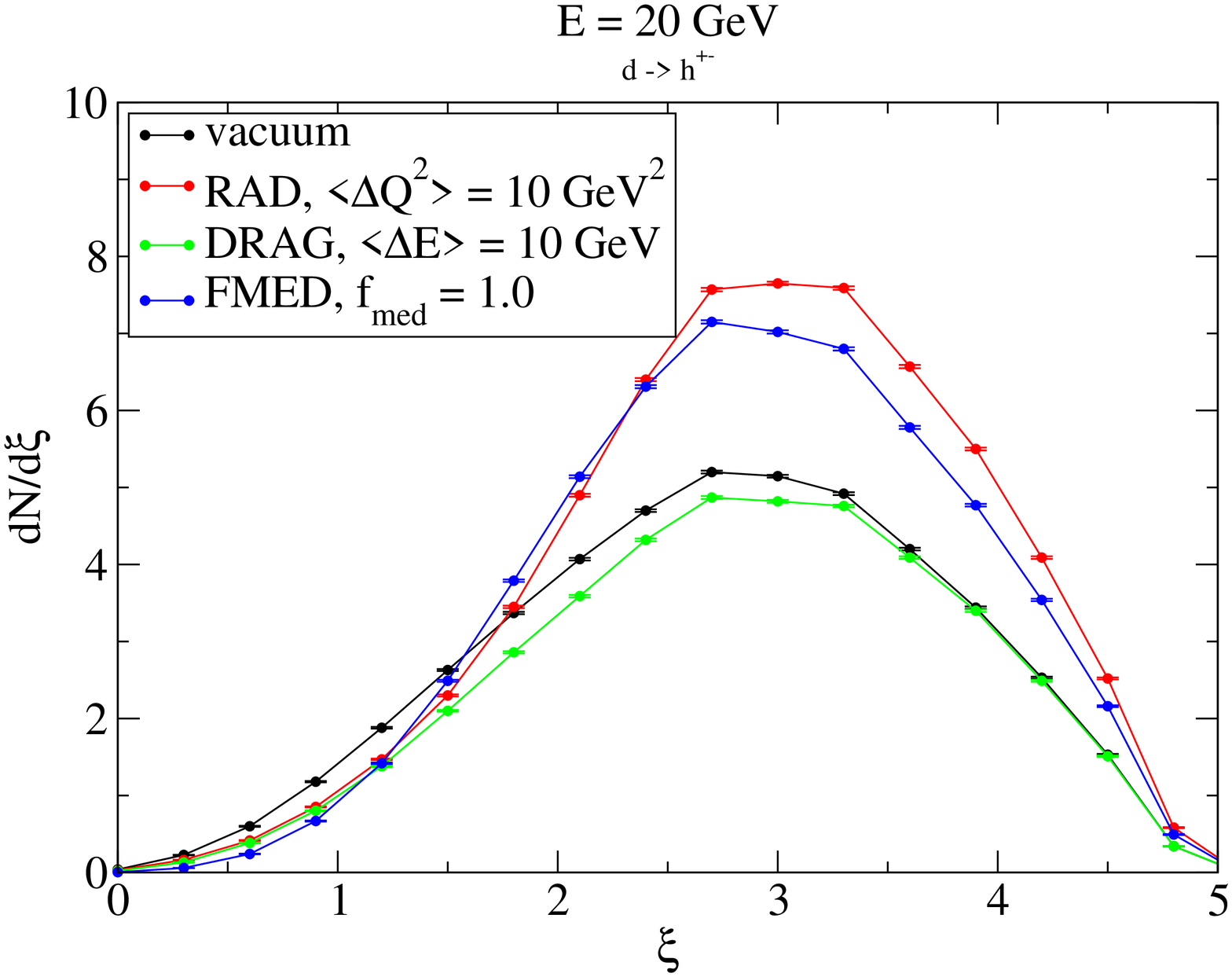, width=6.5cm}{\epsfig{file=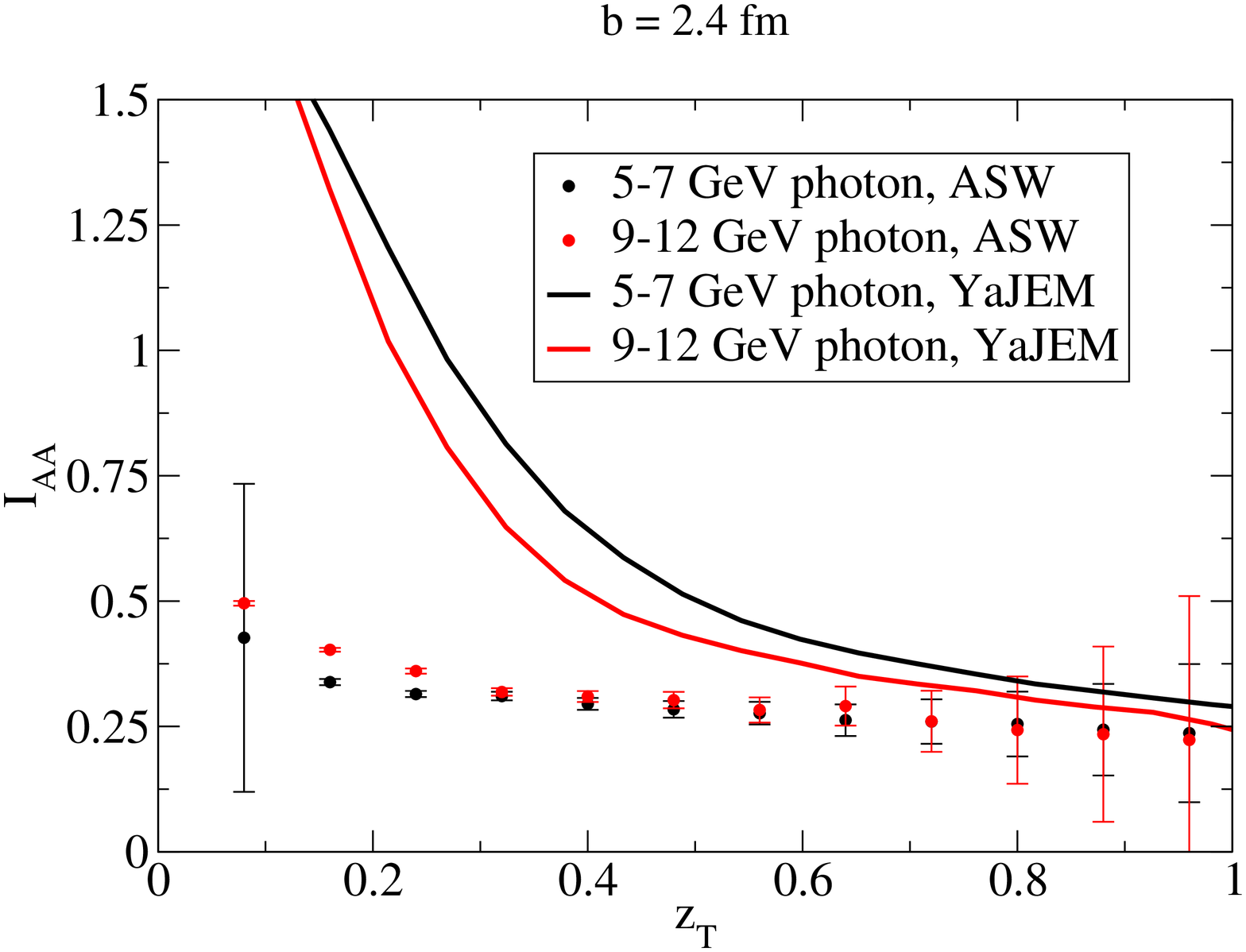, width=6.5cm}}
\caption{\label{F-YaJEM2} Left panel: The 'hump-backed plateau' distribution $dN/d\xi$ obtained in YaJEM for three different scenarios of parton-medium interaction. Right panel: The away side per-trigger yield in Au-Au collisions normalized to the result in p-p collisions $I_{AA}$ in $\gamma$-h correlations obtained in the YaJEM RAD scenario and in a standard leading parton energy loss picture .}
\end{figure}

The right panel of Fig.~\ref{F-YaJEM1} shows the medium-modified fragmentation function in all three different scenarios of parton-medium interaction. All lead to a comparable depletion of the high $z$ region of the fragmentation function, corresponding to leading parton energy loss. To see the modification in the low $z$ part more clearly, it is useful to transform into $\xi = \ln (1/z)$. This is shown in Fig.~\ref{F-YaJEM2}, left panel.

While the DRAG scenario (in which energy flows into the medium) leads to a small suppression in this region, induced radiation is manifest as an enhancement of the plateau. Experimentally, this region in $\xi$ can be probed by $\gamma$-hadron correlations. The expectation  for the away side per trigger yield suppression ratio $I_{AA}$ from YaJEM (with enhanced low $z$ hadron production due to induced radiation) and a standard energy loss calculation (where lost energy is assumed to be redistributed throughout the whole medium) is shown in Fig.~\ref{F-YaJEM2} right panel \cite{gamma-h}. Currently, there is no evidence in the data for a rise of $I_{AA}$ above unity in the data. This might be some evidence that a picture of \emph{perturbative} redistribution of lost energy inside the shower only is not justified and that energy flow in the medium needs to be accounted for.

\begin{figure}[htb]
\epsfig{file=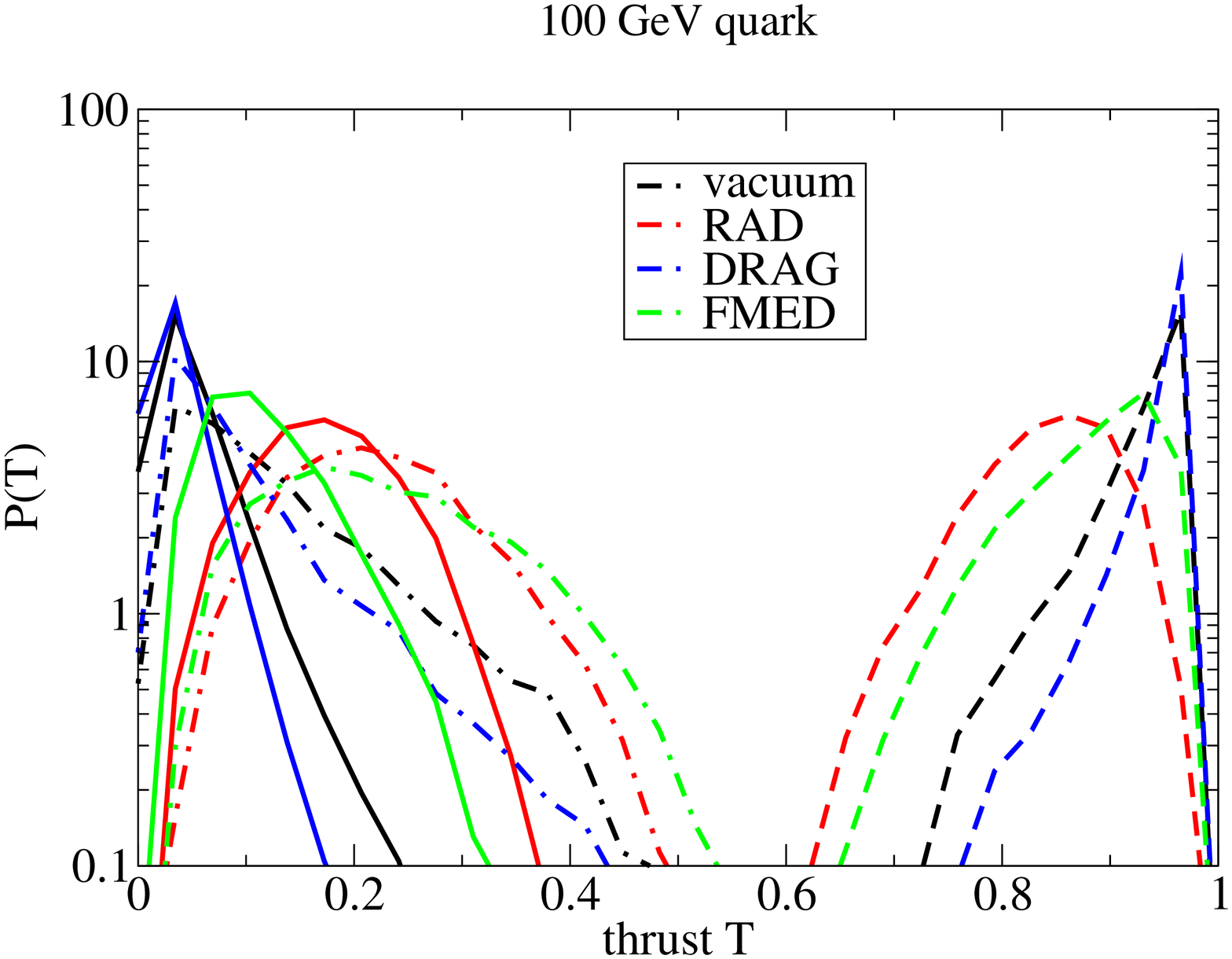, width=6.5cm}{\epsfig{file=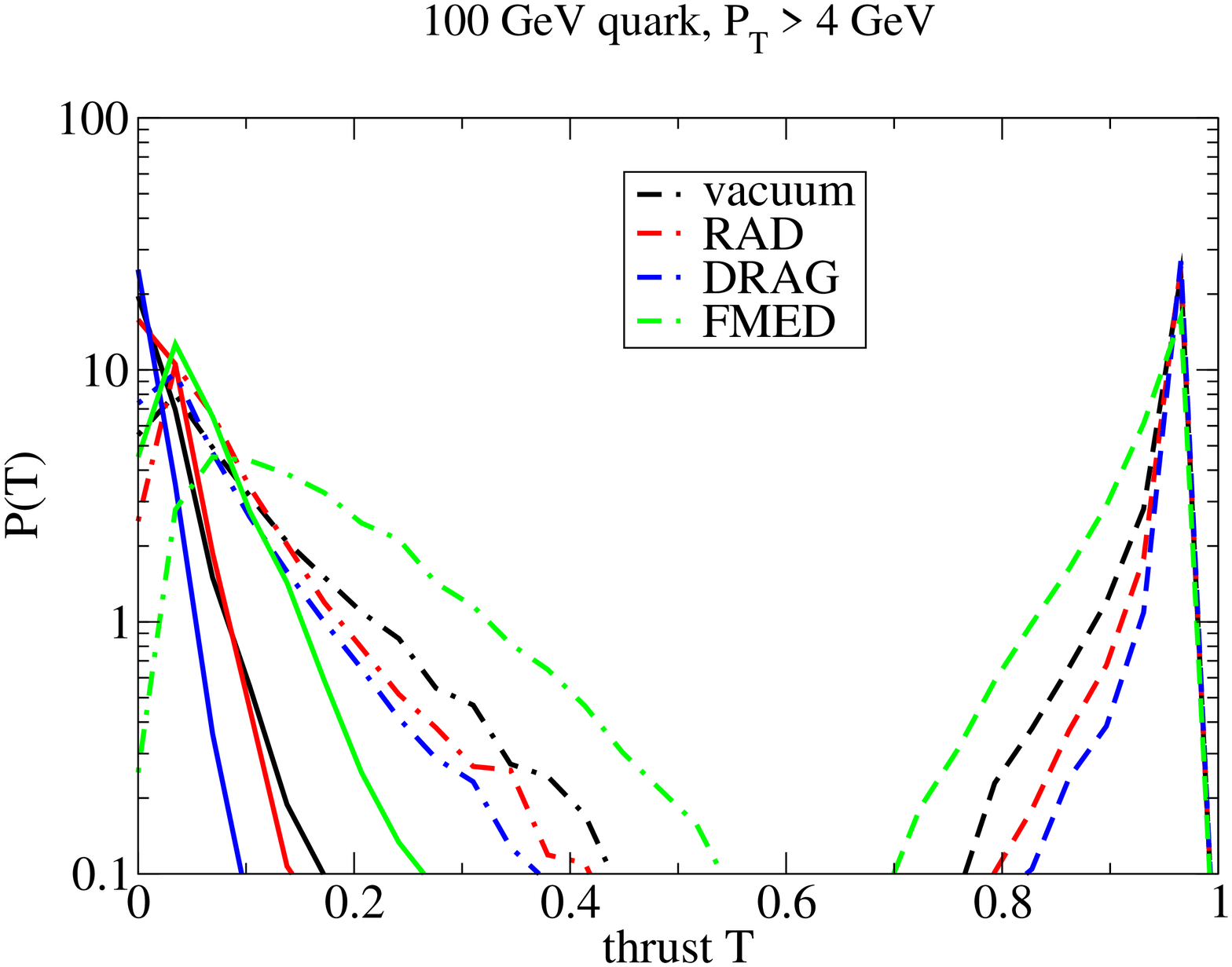, width=6.5cm}}
\caption{\label{F-YaJEM3} Distributions of thrust, thrust major and thrust minor for 100 GeV quark jets in vacuum and for a medium expected in central 5.5 ATeV Pb-Pb collisions at LHC as computed in YaJEM, for all $P_T$ (left) and with a cut of 4 GeV (right).}
\end{figure}

To provide an example for genuine jet observables, Fig.~\ref{F-YaJEM3} shows the distribution of thrust  
$T = \text{max}_{{\bf n}_T} \frac{\sum_i | {\bf p}_i \cdot {\bf n}_T|}{\sum_i|{\bf p}_i|} \quad$, thrust major
$T_{maj} = \text{max}_{{\bf n}_T \cdot {\bf n}=0} \frac{\sum_i | {\bf p}_i \cdot {\bf n}|}{\sum_i|{\bf p}_i|} \quad$ and thrust minor
$T_{min} = \frac{\sum_i | {\bf p}_i \cdot {\bf n}_{mi}|}{\sum_i |{\bf p}_i|}$
for 100 GeV quark jets propagated through medium densities characteristic for LHC Pb-Pb collisions.

There is a general trend that medium-induced radiation makes the event more spherical, regardless if explicit exchange of energy and momentum with the medium is modelled or not, but the $P_T$ dependence of the distribution is very different in both cases. Thus, assuming that the parton shower has zero momentum exchange with the medium may be a bad approximation. This in turn means that the problem of energy redistribution by non-perturbative medium degrees of freedom in all likelihood will have to be dealt with in order to understand medium-modified jets.

\section{Conclusions}

Currently medium-modified shower codes have to be regarded as work in progress. Qualitatively, some of their results look promising. However, when looking into the details, there are often problems in reproducing more differential leading hadron observables. Quantitatively, YaJEM awaits comparison with jet measurements at RHIC, which, due to the complicated jet finding in a heavy-ion background, is not an easy task. There are also some indications that non-perturbative dynamics in the medium might play a role for redistributing the energy and momentum of the initial hard process in addition to the perturbative dynamics of a parton shower. The kinematic range available at the LHC will help much to resolve these questions.

\section*{Acknowledgements}

\end{document}